\documentclass[a4paper]{PoS}
\usepackage{amsmath}
\usepackage{amsfonts}
\usepackage{graphicx}
\usepackage{slashed}
\usepackage{amssymb}

\newcommand{\pbrac}[1]{\left( #1 \right)}
\newcommand{\tbrac}[1]{\left[ #1 \right]}
\newcommand{\cbrac}[1]{\left\{ #1 \right\}}

\title{Phenomenology of the Renormalizable Coloron Model\protect\footnote{This contribution is based on the publications \cite{Chivukula:2013xka,Chivukula:2014rka,Chivukula:2015kua,Chivukula:2011ng,Chivukula:2013xla}.}}

\ShortTitle{Phenomenology of the Renormalizable Coloron Model}

\author{\speaker{Arsham Farzinnia}\\
       Center for Theoretical Physics of the Universe, Institute for Basic Science (IBS)\\Daejeon 305-811, Republic of Korea\\
       E-mail: \email{farzinnia@ibs.re.kr}}

\abstract{The renormalizable coloron model constitutes the minimal extension of the standard model (SM) color sector to $SU(3)_{1c} \times SU(3)_{2c}$, with the spontaneous symmetry breaking of the extended gauge group to the diagonal QCD facilitated by the renormalizable operators. It predicts the existence of the beyond the SM massive color-octet gauge bosons (colorons), colored and uncolored scalar degrees of freedom, as well as potential spectator fermions necessary for anomaly-cancelation purposes. Furthermore, keeping the ordinary chiral quark charge assignments under the extended color gauge group in their most general form, the framework (effectively) captures a large class of models available within the literature. This contribution summarizes the current formal and phenomenological constraints on the free parameter space of the theory, as well as the LHC $\sqrt s = 14$~TeV prospects for discovering its heavy scalar. The model is well-constrained and highly predictive; in particular, it is shown that the parameter space can be thoroughly probed by the LHC and the next generation hadron colliders, making it a promising beyond the SM candidate for exploration. In addition, the significance of the NLO corrections to the coloron production cross section are discussed.}

\FullConference{18th International Conference From the Planck Scale to the Electroweak Scale \\
		 25-29 May 2015\\
		 Ioannina, Greece }

\begin{document}

\section{Introduction}\label{intro}

Colorons are a set of massive color-octet vector bosons which arise in a variety of beyond the standard model (SM) theories, where the SM color gauge group is (effectively) extended to $SU(3)_{1c} \times SU(3)_{2c}$. A spontaneous breaking of this extended color gauge group to its diagonal subgroup $SU(3)_{c}$ (identified as the ordinary QCD containing the usual massless gluons) produces the colorons as a massive copy of the gluon color-octet set,\footnote{This is in analogy with the extensions of the electroweak sector, where the SM $SU(2)_{L} \times U(1)_{Y}$ gauge group is augmented by the additional $SU(2)$ and/or $U(1)$ gauge groups, producing the $W'$ and $Z'$ vector bosons.} but potentially with chiral couplings to quarks. Various types of models (effectively) containing such an extended color gauge group are present within the literature, including: topcolor \cite{Hill:1991at}, the flavor-universal coloron \cite{Chivukula:1996yr}, chiral color (the axigluon) \cite{axigluon}, chiral color with unequal gauge couplings \cite{Martynov:2009en}, flavor non-universal chiral color \cite{Frampton:2009rk}, flavorful top-coloron model \cite{Top-Coloron}, technicolor models with colored technifermions \cite{Chivukula:1995dt}, and extra-dimensional models with KK gluons \cite{KKg}.

In case the spontaneous symmetry breaking in the extended color sector occurs within a renormalizable framework \cite{Chivukula:2013xka,Chivukula:1996yr,recom}, additional colored and uncolored (pseudo-)scalars are predicted within the reach of the LHC. Furthermore, in a formalism where the couplings of the ordinary left- and right-handed quarks to colorons are different (chiral color), potential gauge anomalies may arise, the cancellation of which requires the existence of new chiral spectator fermions with the opposite couplings to colorons as compared with those of the chiral quarks. Hence, the theory introduces a rich spectrum of beyond the SM scalar, vector, and fermionic degrees of freedom within a consistent framework, which may be thoroughly explored by the LHC and the next generation hadron colliders.

\section{Renormalizable Coloron Model}\label{ReCoM}

The spontaneous symmetry breaking in the enlarged color sector is presumed to occur at an energy scale higher than the scale of the electroweak symmetry breaking,
\begin{equation}\label{gaugegroup}
SU(3)_{1c} \times SU(3)_{2c} \times SU(2)_{L} \times U(1)_{Y}  \rightarrow  SU(3)_{c} \times SU(2)_{L} \times U(1)_{Y}  \rightarrow  SU(3)_{c} \times U(1)_{\text{EM}}.
\end{equation}
A renormalizable symmetry breaking framework may be constructed by introducing a complex scalar $\Phi$, which is in the bi-fundamental representation of the two color gauge groups; i.e., it is a $(3,\bar{3})$ under $SU(3)_{1c} \times SU(3)_{2c}$, and singlet under the electroweak gauge group, with the general form
\begin{equation}\label{Phi}
\Phi = \frac{1}{\sqrt{6}} \pbrac{v_{s} + s_{0} + i {\cal A}} {\cal I}_{3\times 3} + \pbrac{G^a_H + i G^a_G}t^a \qquad \pbrac{t^a \equiv \lambda^a/2} \ .
\end{equation}
The $s_{0}$ ($\mathcal A$) is the gauge-singlet scalar (pseudoscalar) component, and $G_{H}^{a}$ is a set of massive color-octet scalars. The color-octet pseudoscalars $G_{G}^{a}$ are the Nambu-Goldstone bosons eaten by the colorons to make the latter massive. The nonzero vacuum expectation value (VEV) of the $s_{0}$ scalar triggers the spontaneous symmetry breaking in the extended color sector.

Defining $\phi$ as the ordinary electroweak Higgs field doublet with the usual components ($v_{h}=246$~GeV),
\begin{equation}\label{phi}
\phi= \frac{1}{\sqrt{2}}
\begin{pmatrix} \sqrt{2}\,\pi^+ \\ v_h+h_0+i\pi^0 \end{pmatrix} \ ,
\end{equation}
one may construct the most general scalar potential composed of $\phi$ and $\Phi$ \cite{Chivukula:2013xka}
\begin{equation}\label{pot}
\begin{split}
V(\phi,\Phi) = &\, \frac{\lambda_h}{6}\pbrac{\phi^\dagger \phi - \frac{v^2_h}{2}}^2 + \lambda_m\pbrac{\phi^\dagger \phi - \frac{v^2_h}{2}} \pbrac{{\rm Tr}\tbrac{\Phi^\dagger \Phi} - \frac{v^2_s}{2}}+ \frac{\lambda_s}{6}\pbrac{{\rm Tr}\tbrac{\Phi^\dagger \Phi}}^2\\
& + \frac{\kappa_s}{2} {\rm Tr}\tbrac{\pbrac{\Phi^\dagger \Phi}^{2}} -\frac{\lambda_s + \kappa_s}{\sqrt{6}}\, v_{s} r_\Delta  \pbrac{{\rm Det}\Phi + {\rm h.c.}}  -\frac{\lambda_s + \kappa_s}{6} \, v_s^{2} \pbrac{1 - r_{\Delta}} {\rm Tr}\tbrac{\Phi^\dagger \Phi} \ ,
\end{split}
\end{equation}
with the potential being bounded from below and its global minimum coinciding with the scalar VEVs for
\begin{equation}\label{stab}
\lambda_h > 0 \ , \qquad \lambda_s^{\prime} > 0 \ , \qquad \kappa_{s} > 0 \ , \qquad \lambda_m^2 < \frac{1}{9} \lambda_h \lambda_s^{\prime} \ , \qquad 0 \le r_{\Delta} \le \frac{3}{2} \ ,
\end{equation}
and $\lambda_{s}^{\prime} \equiv \lambda_{s} + \kappa_{s}$. The scalars $h_{0}$ and $s_{0}$ with nonzero VEVs are mixed with one another due to the $\lambda_{m}$~coupling in the potential \eqref{pot}, and may be diagonalized using an orthogonal rotation
\begin{equation}\label{massbasis}
\begin{pmatrix} h_0\\ s_0 \end{pmatrix}
=  \begin{pmatrix} \cos\chi & \sin\chi \\ -\sin\chi & \cos\chi \end{pmatrix} \begin{pmatrix} h \\ s \end{pmatrix} \ , \quad \cot 2\chi \equiv \frac{1}{6\lambda_m} \tbrac{ \lambda_s^{\prime}\pbrac{1-\frac{r_{\Delta}}{2}} \frac{v_s}{v_h} - \lambda_h \frac{v_h}{v_s} } \ .
\end{equation}
Hence, the model contains two physical Higgs-like scalars, $h$~and~$s$. Subsequently, one obtains the physical scalar masses $m_{h} = 125$~GeV \cite{LHCHiggs}, $m_{s}$, $m_{\cal A}$, and $m_{G_{H}}$, satisfying the relations
\begin{equation}\label{massrel}
m_{\mathcal A}^{2} \leq \frac{3}{2} \, m_{G_{H}}^{2} \ , \qquad m_{\mathcal A}^2 \leq 9\tbrac{m_h^2 \sin^2 \chi + m_s^2 \cos^2\chi} \ .
\end{equation}
Tevatron searches exclude scalar color-octet masses within the $50 \lesssim m_{G_{H}} \lesssim 125$~GeV range \cite{Aaltonen:2013hya}; hence, we assume $m_{G_{H}} > m_{h}=125$~GeV.

The coloron mass can be expressed in terms of the two color gauge group couplings, $g_{s_{1}}$ and $g_{s_{2}}$, and the singlet VEV according to \cite{Chivukula:2011ng,Chivukula:2013xla}
\begin{equation}\label{MC}
M_C = \sqrt{\frac{2}{3}}\frac{g_s\, v_{s}}{\sin 2\theta_c} \ , \qquad
\sin\theta_c \equiv \frac{g_{s_1}}{\sqrt{g_{s_1}^2+g_{s_2}^2}}\ ,
\end{equation}
whereas, the ordinary QCD strong coupling is given by
\begin{equation}\label{gs}
\frac{1}{g_s^2} = \frac{1}{g_{s_1}^2}+\frac{1}{g_{s_2}^2}\ .
\end{equation}
Experimentally, the lower bound on the coloron mass is determined to lie within the TeV ballpark by the Tevatron and LHC searches \cite{ColoronLim}.

In general, the chiral eigenstates of each quark flavor may be charged under different $SU(3)_{i\,c}$ color gauge groups ($i=1,2$), which results in the chiral couplings of the quarks to colorons\footnote{The quark couplings to gluons remain vector-like, regardless of the chiral charge assignment.}
\begin{equation} \label{Lferm}
\begin{split}
&\mathcal{L}_{\rm quark} = \bar{q}^i i\left[\slashed{\partial}-i g_s \slashed{G}^a t^a -i \slashed{C}^a t^a \left(g_L P_L+g_R P_R\right)\right]q_i \ , \\
&P_{L,R}\equiv \frac{1\mp \gamma_5}{2}\ , \quad  g_L, g_R \in \left \{ -g_s \tan \theta_c, g_s \cot \theta_c \right \} \ .
\end{split}
\end{equation}
This, however, may give rise to potential gauge anomalies in the extended color sector, endangering the consistency of the theory. Such potential anomalies may be canceled by introducing an appropriate number of additional chiral spectator fermions, which have the opposite chirality charges under the extended color gauge group with respect to the ordinary quarks \cite{Chivukula:2013xla}. Both the left- and the right-handed spectator generations are assumed to be doublets under the $SU(2)_{L}$ gauge group, possessing a $U(1)_{Y}$ hypercharge $+1/6$ \cite{Chivukula:2015kua}. Hence, the electric charges of the spectator fermions resemble those of their corresponding quark counterparts ($+2/3$ for the up-like spectator and $-1/3$ for the down-like spectator). The (flavor-universal) spectator fermion masses are generated via their Yukawa interactions with the complex scalar $\Phi$ \eqref{Phi},
\begin{equation}\label{LSpec}
- y_{Q} \tbrac{\bar{Q}^{k}_{R} \, \Phi\, Q^{k}_{L} + \bar{Q}^{k}_{L} \, \Phi^\dagger \, Q^{k}_{R}} \ , \qquad M_{Q} = \frac{y_{Q}}{\sqrt 6}\, v_{s} \ ,
\end{equation}
where, $Q^k_{L(R)}$ denotes a left(right)-handed spectator doublet, and $k$ is the generation index. The lower bound on the spectator fermion masses also resides within the TeV region \cite{Spect}.

Assuming the aforementioned spectator fermion properties, the described minimal framework contains additionally eight free parameters, which may be taken as the following set of the physically relevant quantities \cite{Chivukula:2013xka,Chivukula:2014rka}:
\begin{equation}\label{freepar}
\cbrac{v_{s}, \sin \chi, m_{s}, m_{\cal A}, m_{G_{H}}, M_{C}, M_{Q}, N_Q} \ .
\end{equation}
Here, $N_{Q}$ represents the number of spectator fermion generations required for the anomaly-cancelation purposes, depending on the original assignment of the quark chirality charges within the extended color sector.

\section{Formal and Phenomenological Constraints}\label{constr}

Having reviewed the formalism of the renormalizable coloron model in the previous section, let us discuss the constraints on the model's free parameter space \eqref{freepar} arising from various theoretical and experimental considerations. On the theoretical side, the parameter space can be constrained by demanding the tree-level potential \eqref{pot} to be bounded from below with its global minimum coinciding with the scalar VEVs $v_{h}$ and $v_{s}$ (c.f. \eqref{stab}), as well as imposing the unitarity condition by performing a full coupled-channel analysis of the two-body scalar scatterings. These analyses have been presented in detail in \cite{Chivukula:2013xka}. Furthermore, a thorough study of the vacuum stability and triviality of the model has been performed in \cite{Chivukula:2015kua}, and the viable regions of the parameter space are identified by imposing the conditions \eqref{stab} on the running couplings and, at the same time, requiring the absence of any Landau poles up to a cutoff energy of 100~TeV.

Experimentally, the data from the electroweak precision tests can be utilized to constrain the $s$~boson mass as a function of the scalar mixing angle, since this second Higgs-like scalar is also capable of interacting with the electroweak gauge bosons. In addition, direct measurements of the 125~GeV $h$~boson couplings by the LHC confine the values of the scalar mixing angle. The results of these experimental considerations at 95\%~C.L. have also been discussed in detail in \cite{Chivukula:2013xka}.

\begin{figure}
\includegraphics[width=.33\textwidth]{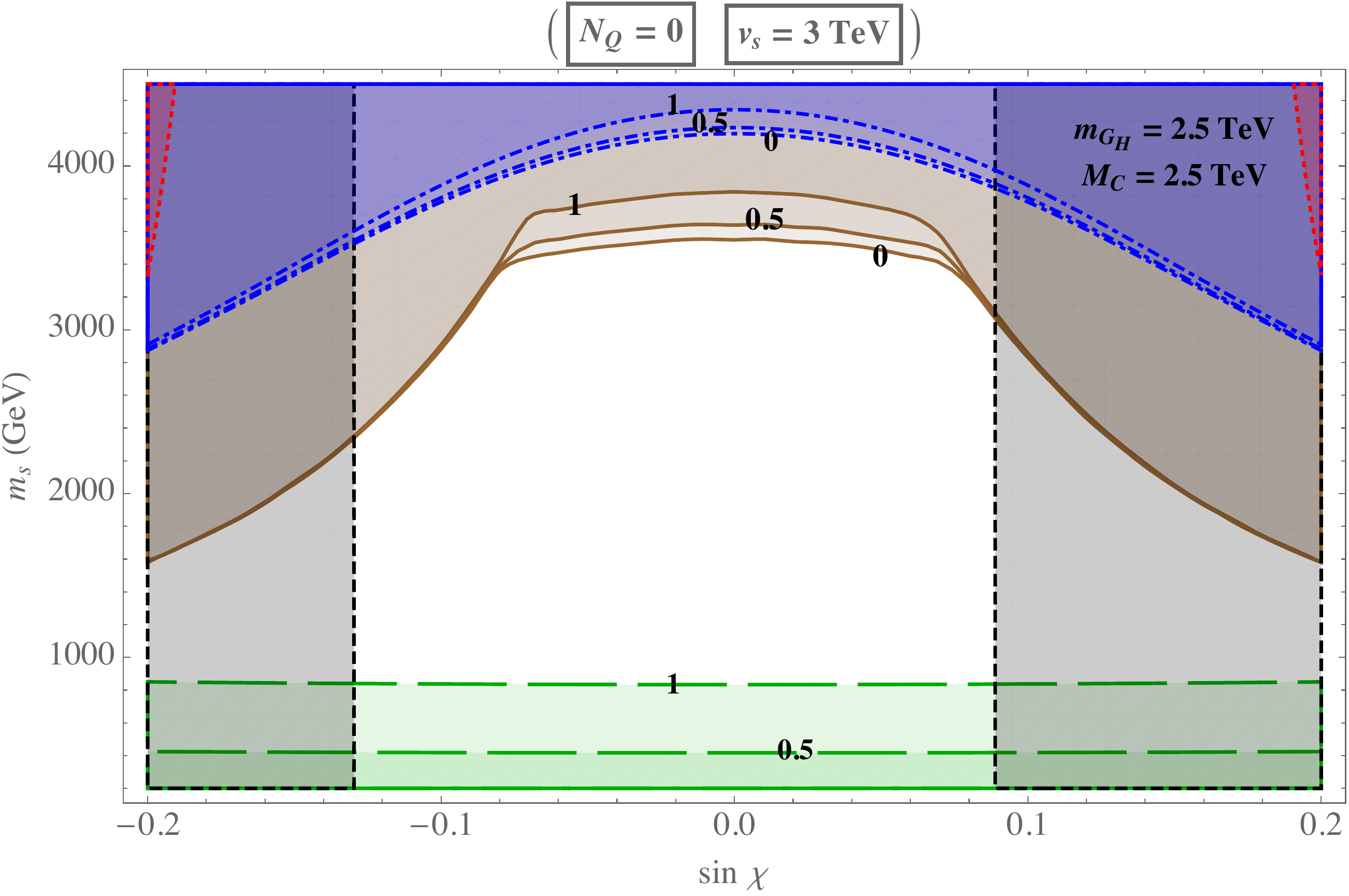}
\includegraphics[width=.33\textwidth]{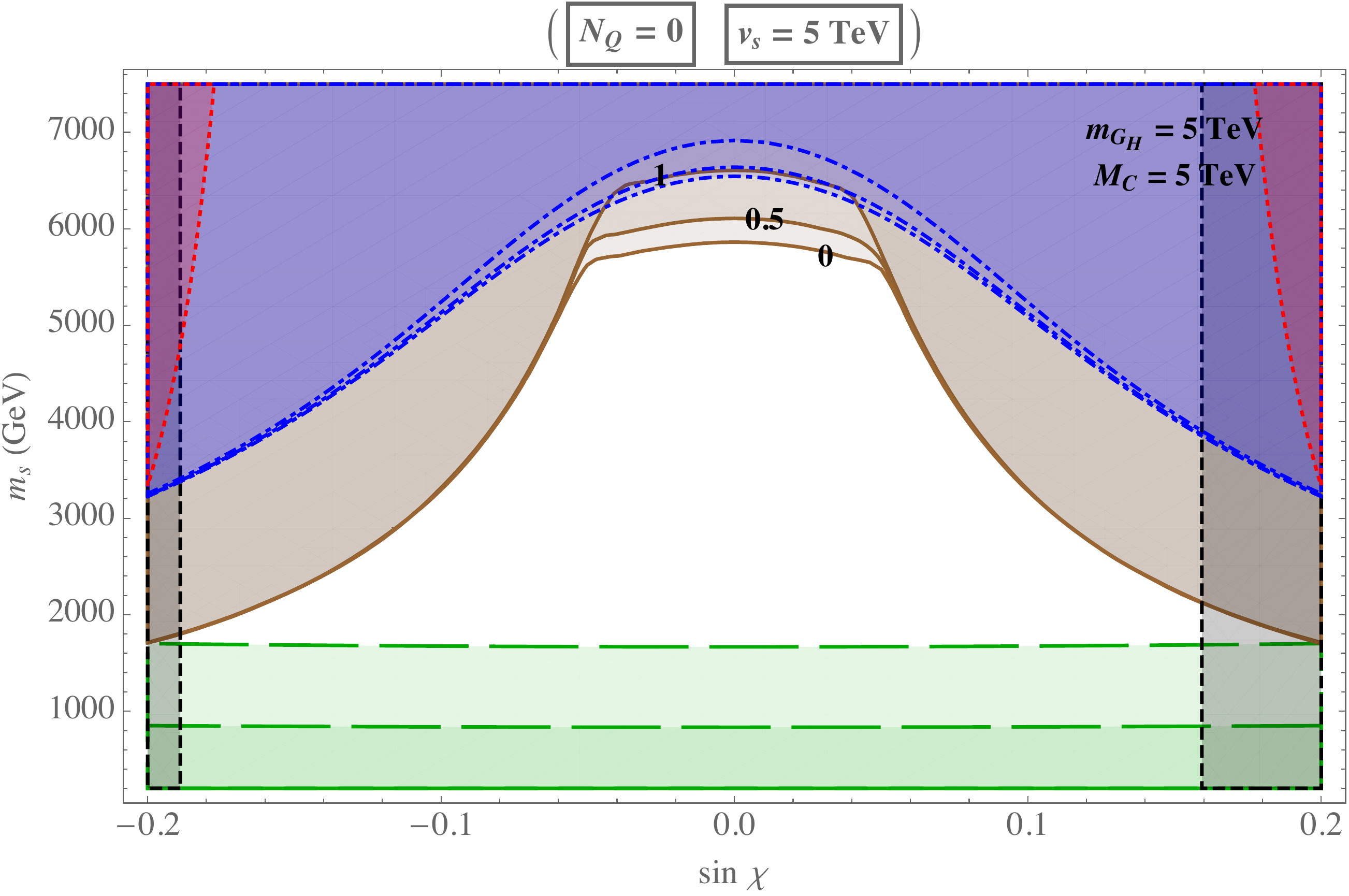}
\includegraphics[width=.33\textwidth]{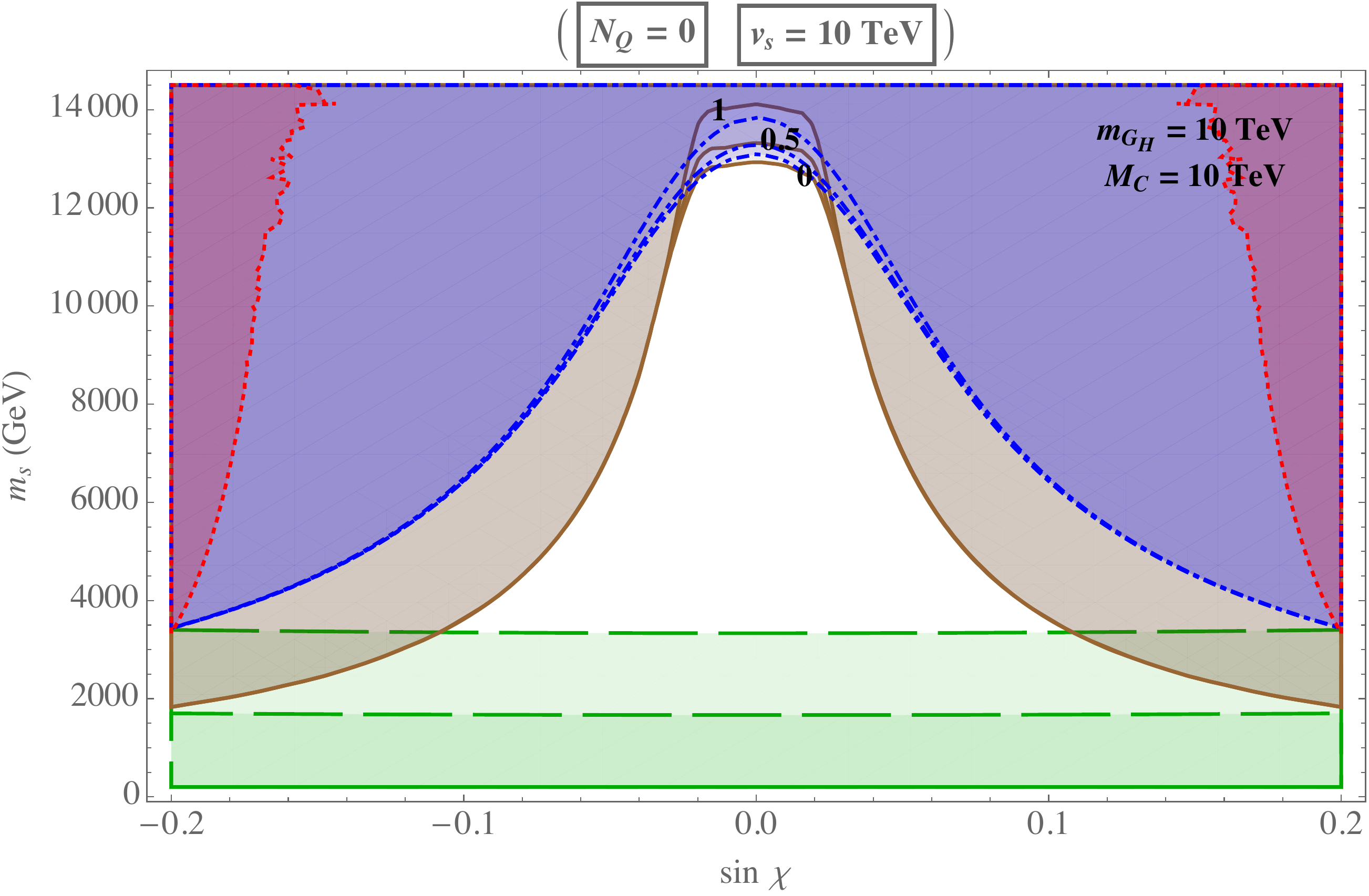}
\includegraphics[width=.33\textwidth]{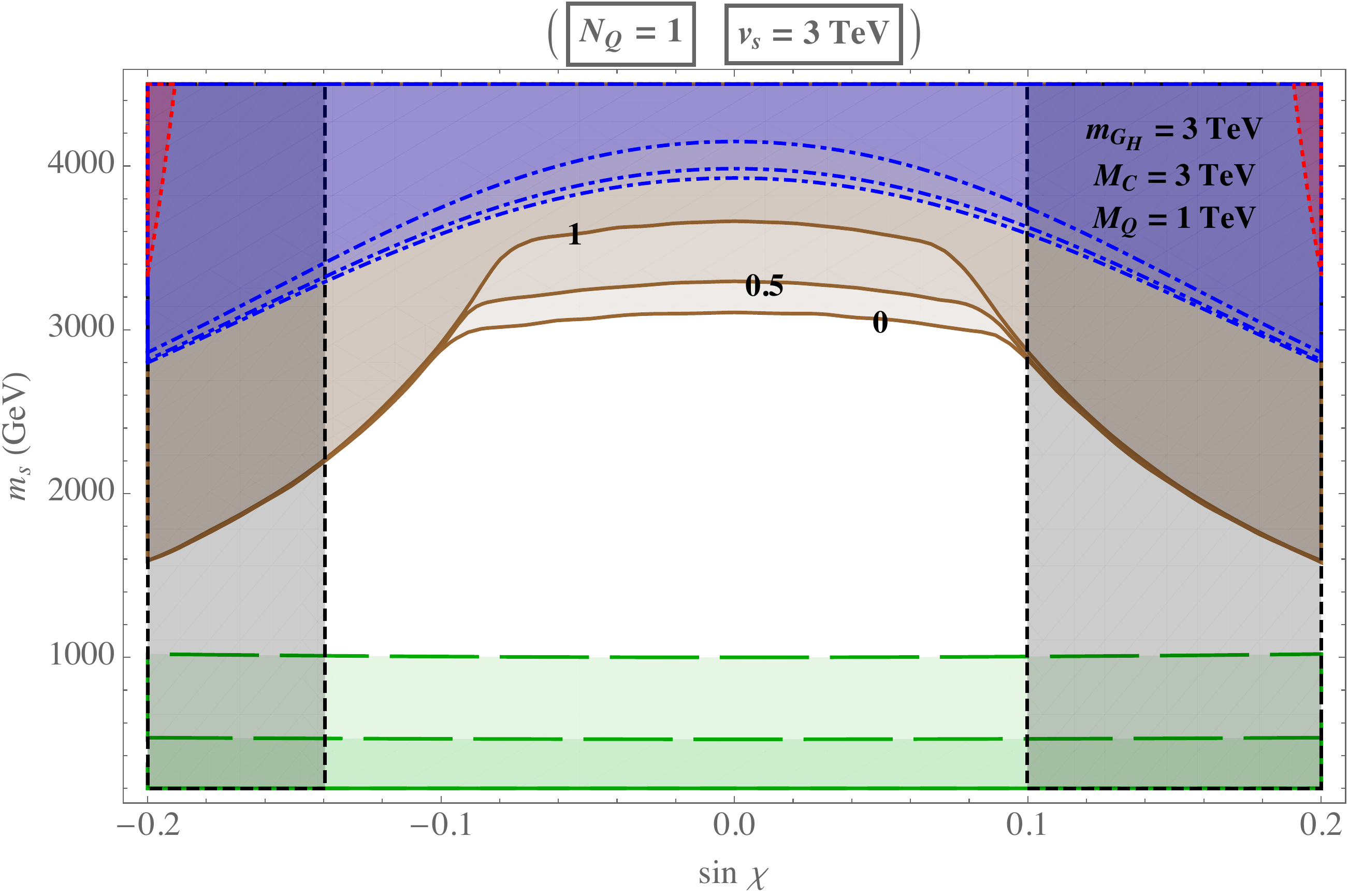}
\includegraphics[width=.33\textwidth]{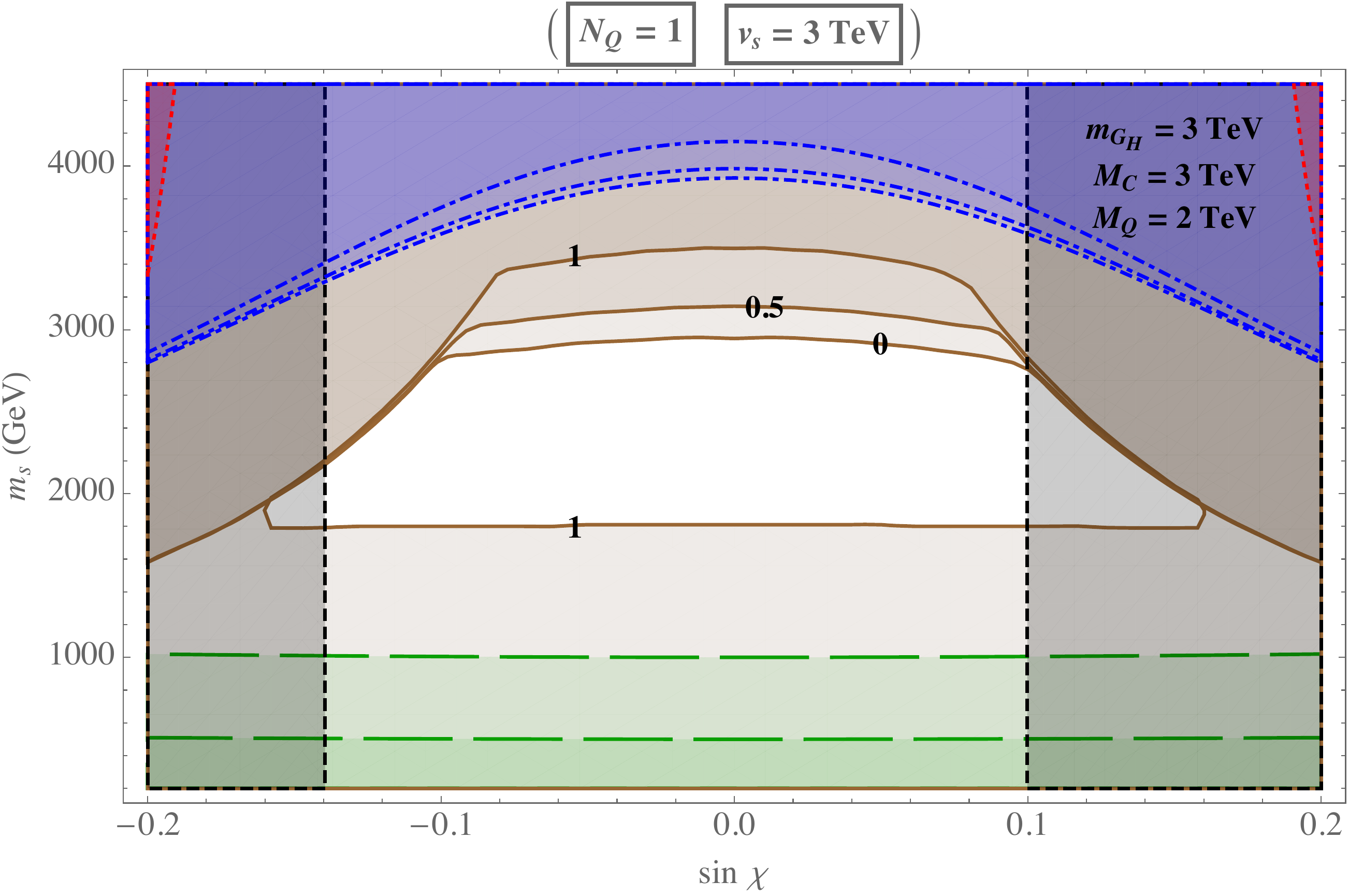}
\includegraphics[width=.33\textwidth]{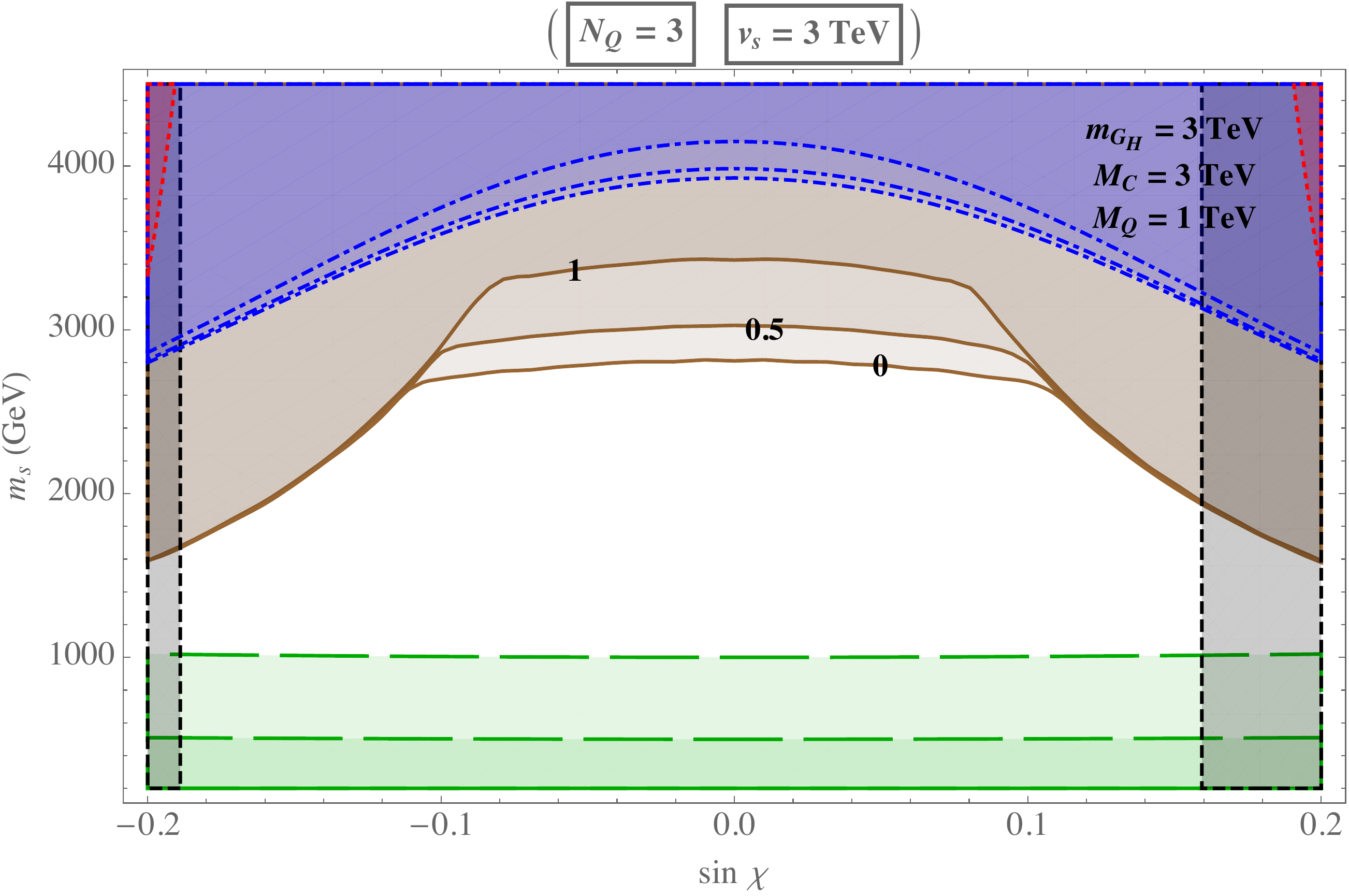}
\caption{The $\sin\chi - m_s$ exclusion plots for various benchmark values of the remaining free parameters of the model, displaying the constraints from imposing the global minimum of the potential coinciding with the VEVs (long-dashed green line), unitarity (dot-dashed blue line), and vacuum stability and triviality up to 100~TeV (solid brown line), as well as the electroweak precision tests (dotted red line), and direct Higgs couplings' measurements by the LHC (short-dashed black line). The enumerated curves correspond to several values of $n \equiv m_{\mathcal A}/m_{G_H}$ between 0 and 1. All colored regions are excluded. (plots taken from \cite{Chivukula:2015kua})}
\label{formalandpheno}
\end{figure}

The aforementioned constraints are summarized within the exclusion plots presented in Fig.~\ref{formalandpheno} for various benchmark values of the model's free parameters. One observes that the theoretical considerations generally confine the mass range of the $s$~boson; whereas, the experimental data constrain the scalar mixing angle. Moreover, heavier spectator fermions are severely disfavored, due to their destabilizing effect on the vacuum of the potential.

\begin{figure}
\begin{center}
\includegraphics[width=.4\textwidth]{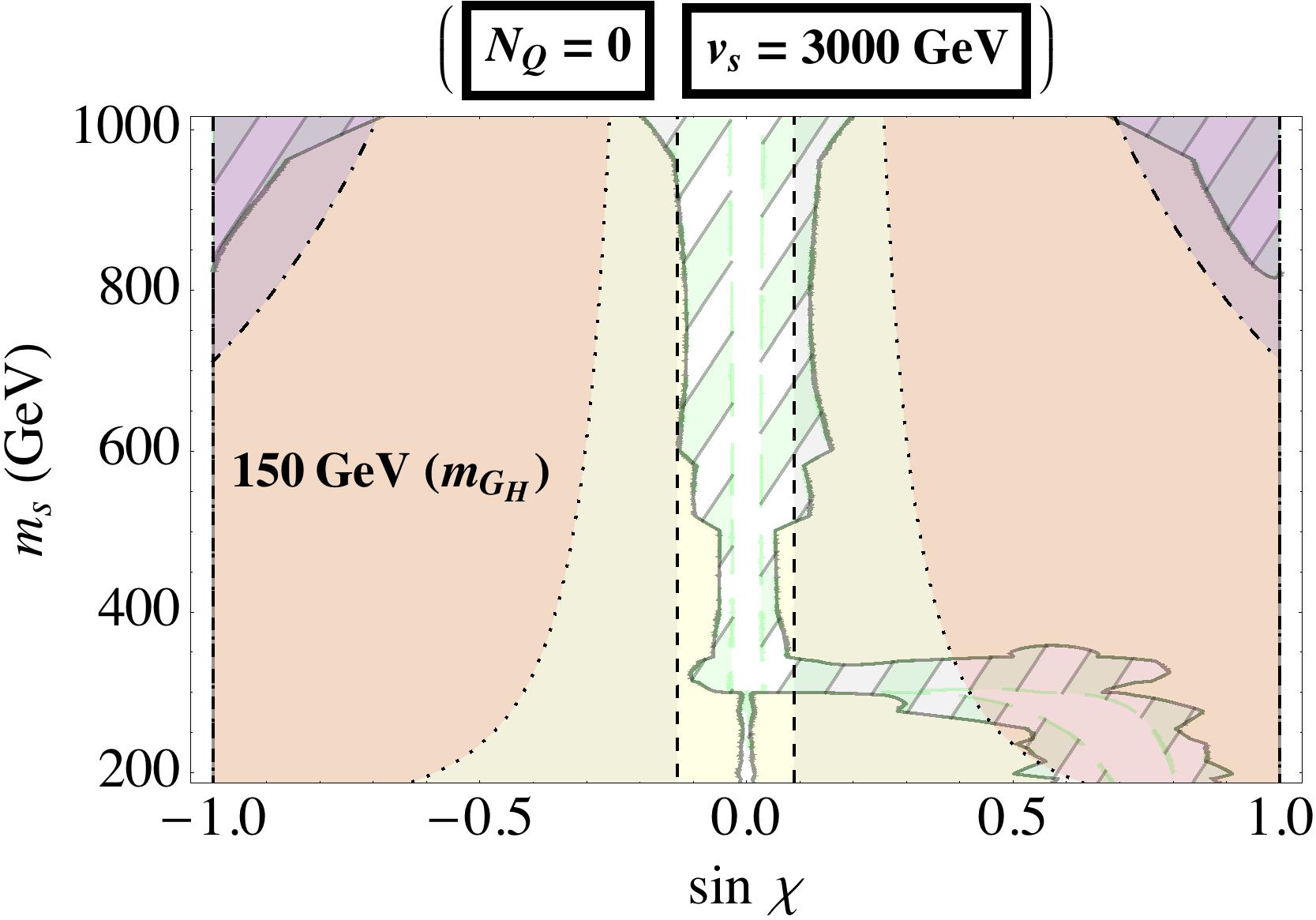}
\includegraphics[width=.4\textwidth]{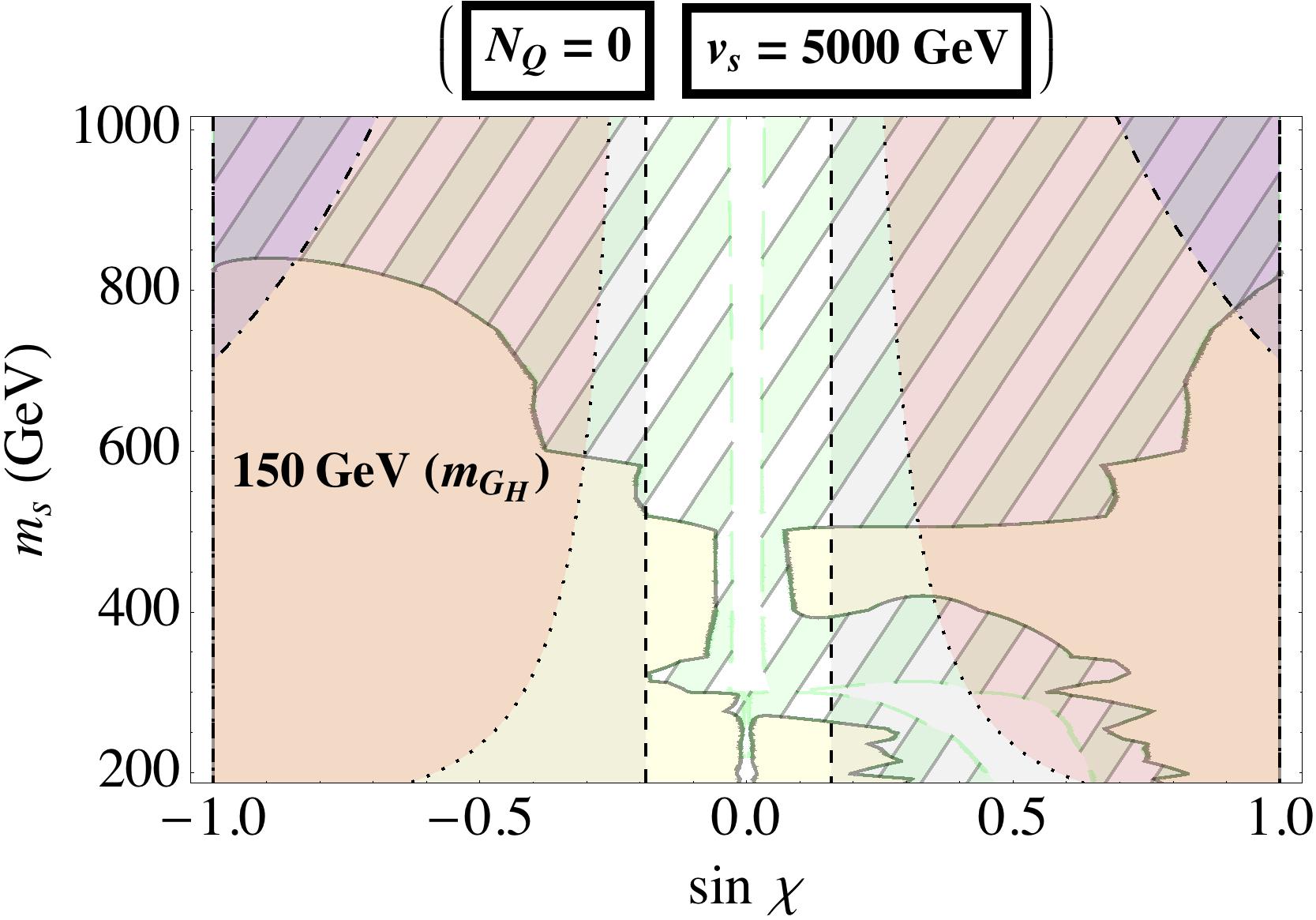}
\includegraphics[width=.4\textwidth]{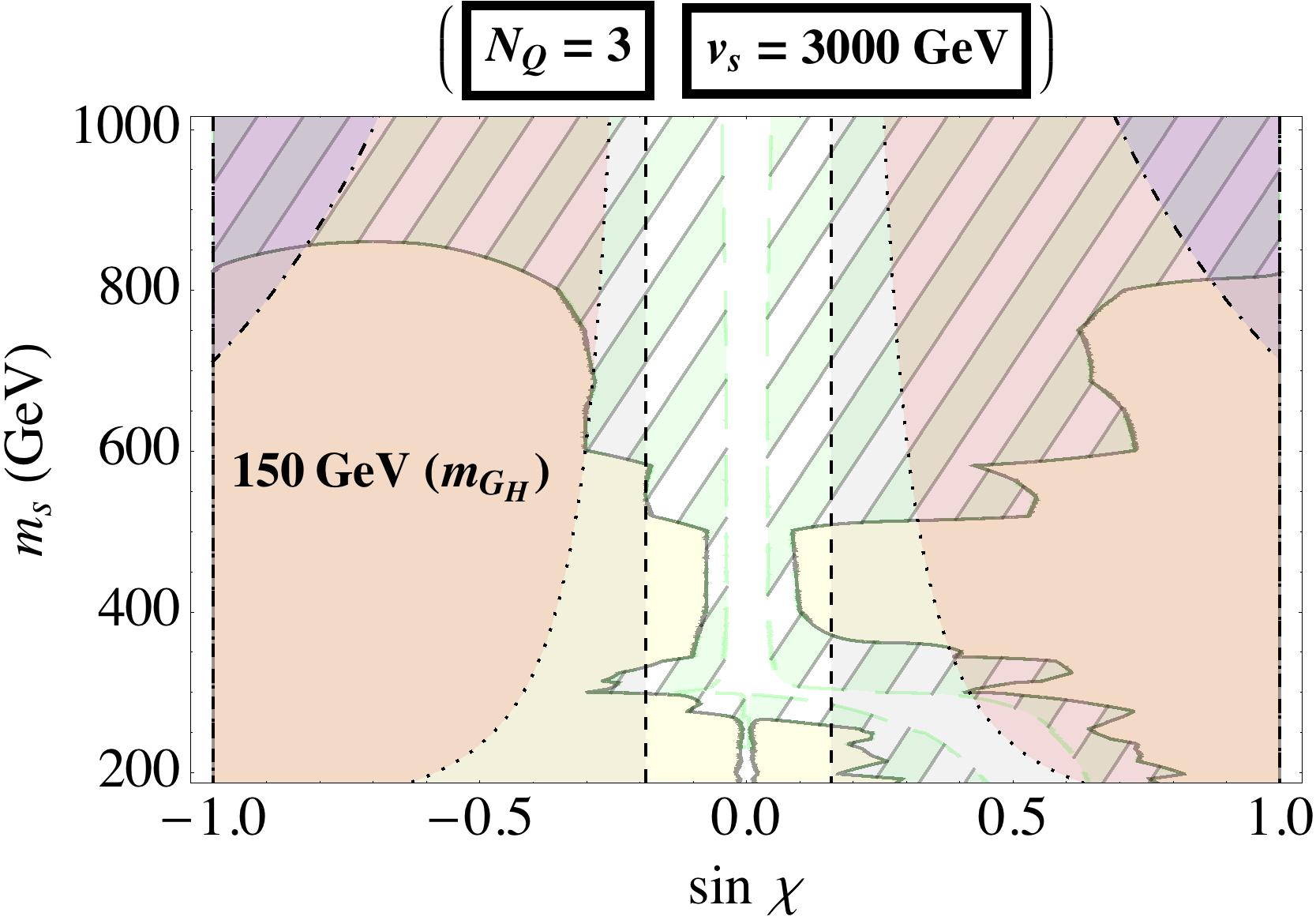}
\includegraphics[width=.4\textwidth]{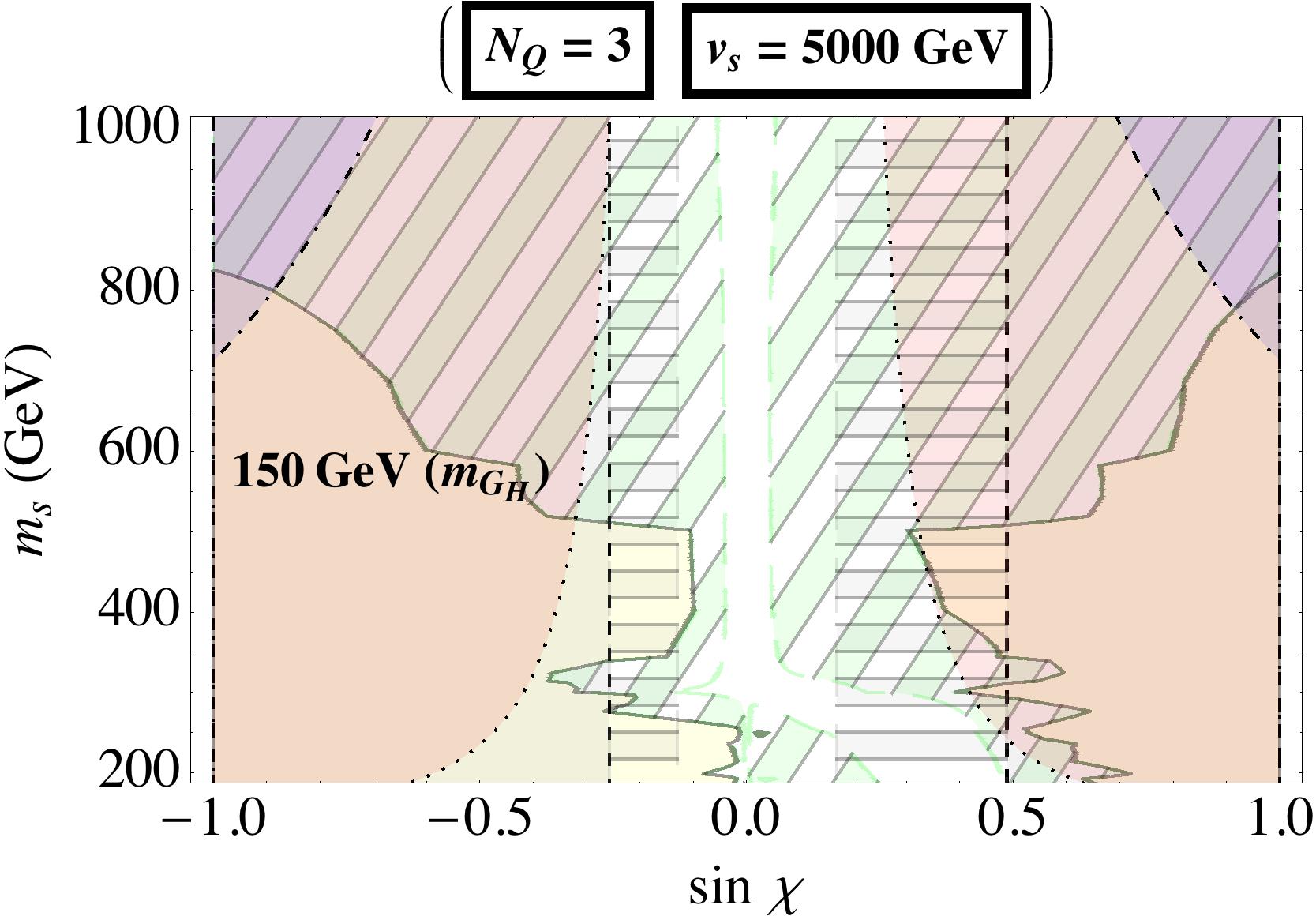}
\caption{The $\sin\chi - m_s$ exclusion plots for various benchmark values of the remaining free parameters of the model, covering the heavy $s$~scalar mass range $200 \leq m_s \leq 1000$~GeV and the full range of mixing angle values $-1\leq \sin \chi \leq 1$. The displayed constraints arise from unitarity (dot-dashed purple), electroweak precision tests (dotted orange), direct Higgs couplings' measurements by the LHC (vertical dashed gray), and the $\sqrt s = 7,8$~TeV LHC heavy $s$~boson searches (solid yellow). The inclined-stripped green region corresponds to the $s$~boson exclusion projections for $\sqrt s = 14$~TeV ATLAS with an integrated luminosity of 300 fb$^{-1}$. The horizontally-stripped gray region in the bottom-right plot indicates the projected exclusion by more precise measurements of the 125~GeV $h$~boson. Universal values of the pseudoscalar and scalar color-octet masses, $m_\mathcal{A} = m_{G_{H}}= 150$~GeV, are selected for the purpose of illustration. The dependence on the coloron and spectator masses, $M_C$ and $M_Q$, is negligible. (plots taken from \cite{Chivukula:2014rka})}
\label{sproj}
\end{center}
\end{figure}

For the $s$~boson mass range $200 \leq m_s \leq 1000$~GeV, competitive bounds at 95\%~C.L. can be obtained from the LHC heavy Higgs searches \cite{HeavyH}. In addition, the ATLAS prospects for discovering the heavy $s$~boson \cite{ATLASHeavyS}, as well as its projections for the more precise measurements of the 125~GeV $h$~boson properties \cite{ATLASHproj}, with $\sqrt s = 14$~TeV and an integrated luminosity of 300 fb$^{-1}$ may be utilized to probe the anticipated LHC reach with regards to the renormalizable coloron model's parameter space. These considerations have been analyzed in \cite{Chivukula:2014rka}, and the results are summarized within the exclusion plots depicted in Fig.~\ref{sproj} for various benchmark values of the free parameters, also incorporating some of the previously mentioned constraints for comparison. It is evident that virtually the entire parameter space of this theory can be probed by the LHC, making it a promising beyond the SM candidate for exploration.\footnote{This is perhaps not surprising, given the fact that the model comprises the minimal extension of the color sector, while the LHC, as a hadron collider, is intrinsically the ideal environment for exploring such strong interactions.} 

\section{Hadron Collider Coloron Production at Next-to-Leading Order}\label{NLO}

\begin{figure}
\includegraphics[width=.33\textwidth]{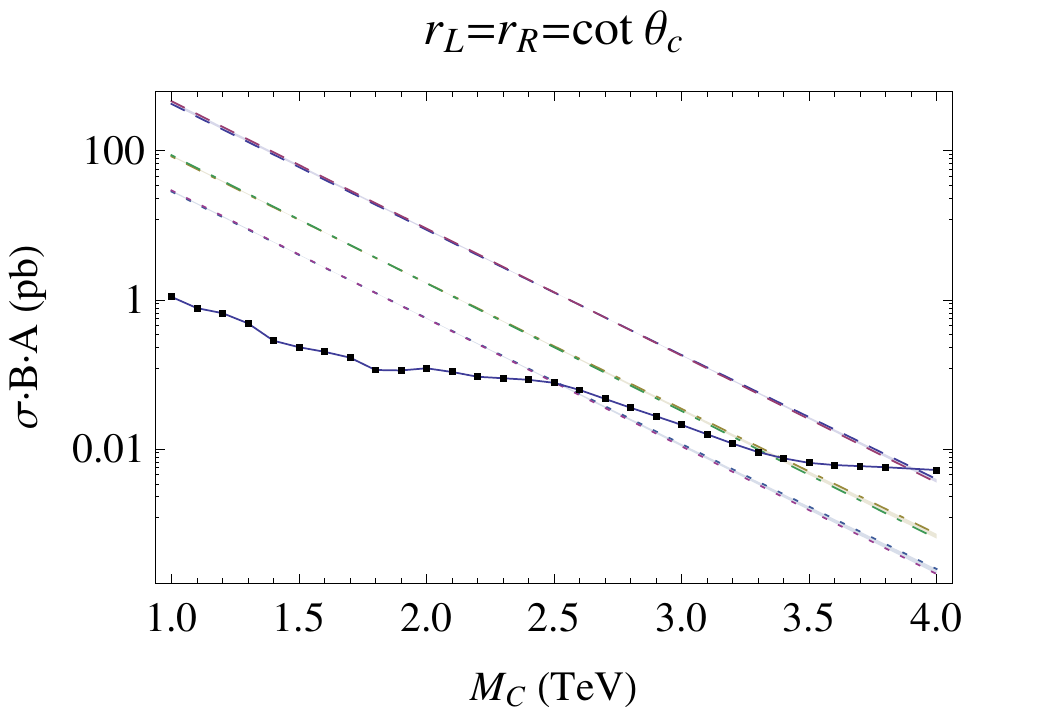}
\includegraphics[width=.33\textwidth]{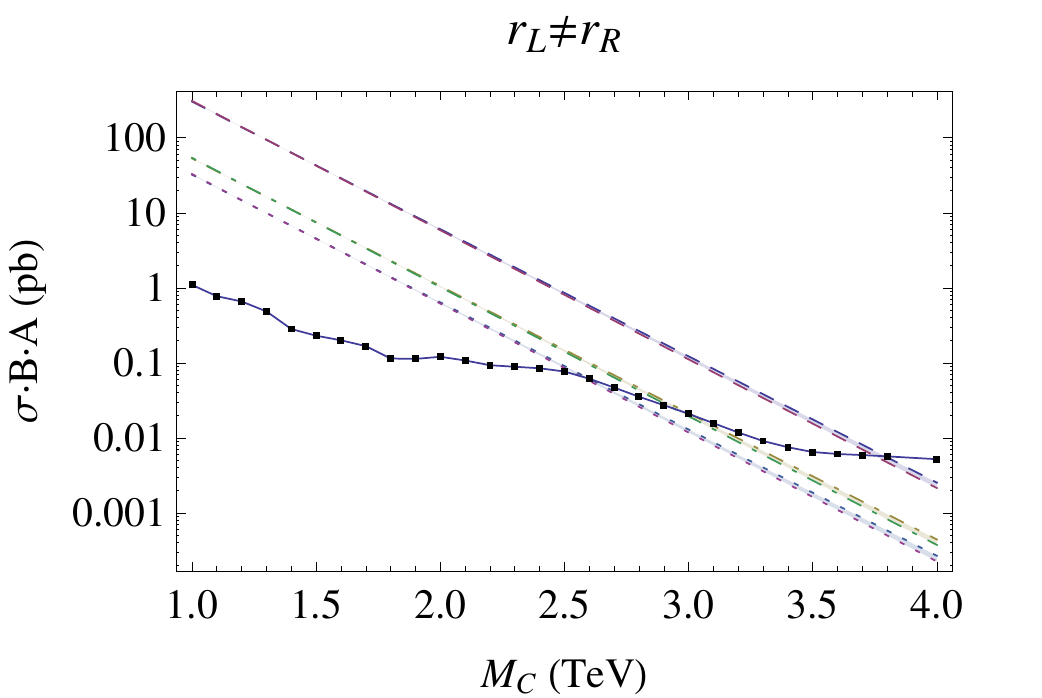}
\includegraphics[width=.33\textwidth]{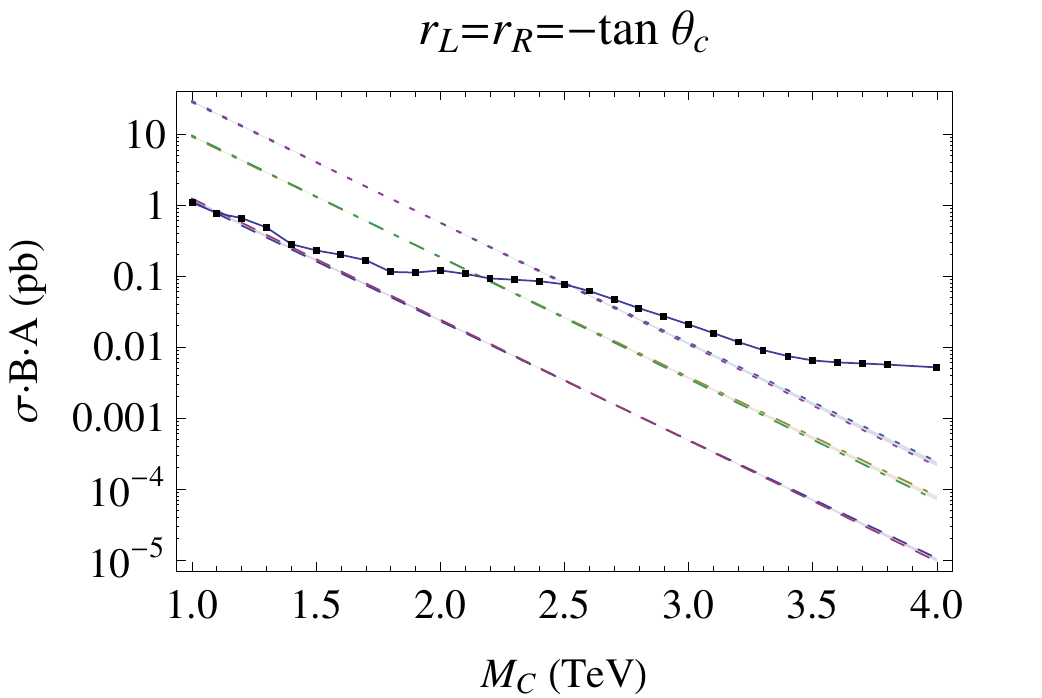}
\caption{On-shell coloron NLO production cross section times branching-ratio to quarks at the LHC ($\sqrt{s}=$ 7 TeV), corrected for the detector acceptance. Three possible flavor-universal scenarios for the quark charge assignments are considered, and within each plot three benchmark values of the gauge mixing angle are shown: $\sin^2\theta_c\vert_{\mu=M_C}=$ 0.05 (dashed), 0.25 (dot-dashed), and 0.5 (dotted). The cross section curves are plotted for the factorization scale, $\mu_{F}$, ranging from $M_C/2$ to $2\,M_C$, and the thickness of each curve reflects the weak dependence of the NLO cross section on the factorization scale. The CMS \protect\cite{CMS} upper limit (solid line) on the cross-section times dijet branching ratio for a narrow resonance is also plotted. The axigluon \protect\cite{axigluon} corresponds to the middle $r_L \neq r_R$ plot with $\sin^2\theta=0.5$; therefore, a narrow axigluon resonance is constrained to have a mass larger than 2.6 TeV. (plots taken from \cite{Chivukula:2011ng})}
\label{CS}
\end{figure}

The first complete calculation of the coloron production cross section at the next-to-leading order (NLO) within a hadron collider has been presented in \cite{Chivukula:2011ng,Chivukula:2013xla},\footnote{The NLO calculation is performed using the (non-renormalizable) non-linear sigma model for the gauge symmetry breaking in the extended color sector. Nevertheless, the virtual corrections due to the additional scalar degrees of freedom present within the formalism of the renormalizable coloron model are not anticipated to substantially affect the obtained results, since the NLO corrections are dominated by the real emission processes in a hadron collider (in particular, a gluon and a (anti)quark in the initial state with a coloron and a (anti)quark in the final state).} taking into account the virtual corrections, as well as the soft and collinear real emission processes. The leading-order coloron production proceeds through the tree-level quark-antiquark pair annihilation. The virtual corrections at one-loop consist of the corrections to the quark self-energy, the vertex, as well as the coloron (mixed) vacuum polarization amplitude. The real emission processes take into account the emission of a soft/collinear gluon from the (anti)quark or the coloron, and additionally the Compton scattering, with a gluon and a (anti)quark in the initial state producing a coloron and a collinear (anti)quark in the final state. Moreover, colorons can be produced at one-loop via the gluon-fusion process \cite{Chivukula:2013xla}, which is forbidden at the tree-level.\footnote{The Landau-Yang's theorem, prohibiting the coupling of a massive spin-1 state to two massless spin-1 states to all orders in perturbation theory, does not apply to non-Abelian gauge theories.}

The full NLO production cross section at a hadron collider can be expressed as the convolution of the NLO partonic cross sections with the quarks and gluon parton distribution functions (PDF) inside each colliding proton
\begin{equation}
\label{eq:barePDFcs}
\begin{split}
\sigma^{\text{NLO}} =&\, \int d x_1 \int d x_2 \bigg\{
\sum_q\tbrac{f^0_q(x_1) f^0_{\bar{q}}(x_2) + f^0_{\bar{q}}(x_1) f^0_q(x_2)}\pbrac{\hat{\sigma}_{q\bar{q}\to C}^{(0)}
+\hat{\sigma}_{q\bar{q}\to C}^{(1)} + \hat{\sigma}_{q\bar{q}\to g C}^{(1)}}  \\
&+ \sum_q\tbrac{f^0_q(x_1) f^0_g(x_2)
+ f^0_g(x_1) f^0_q(x_2)+ f^0_{\bar{q}}(x_1) f^0_g(x_2) + f^0_g(x_1) f^0_{\bar{q}}(x_2)} \hat{\sigma}_{q g\to q C}^{(1)} \\
&+f^0_g(x_1) f^0_g(x_2)\ \hat{\sigma}_{gg\to C}^{(1)}
\bigg\} \ ,
\end{split}
\end{equation}
where, $\hat{\sigma}_{q\bar{q}\to C}^{(0)}$ represents the tree-level partonic cross section, the $\hat{\sigma}^{(1)}$ terms correspond to the various NLO contributions, and $f^0$ terms are the bare PDFs. It has been shown in \cite{Chivukula:2011ng} that the UV divergences of \eqref{eq:barePDFcs} can be subtracted using the two gauge-invariant counterterms of the theory (corresponding to the renormalization of the original gauge couplings, $g_{s_{1}}$ and $g_{s_{2}}$), and all the IR divergences cancel between the virtual correction and the real-emission cross sections, as expected, except for some collinear singularities proportional to the Altarelli-Parisi evolutions. The latter renormalize the bare PDFs. Furthermore, the production cross section via the gluon-fusion process has been demonstrated to be negligible as compared with the remaining NLO contributions even at $\sqrt s = 14$~TeV \cite{Chivukula:2013xla}.

Fig.~\ref{CS} exhibits the NLO production cross section \eqref{eq:barePDFcs} for on-shell colorons, multiplied by the decay branching-ratio to quarks and the detector acceptance as a function of the coloron mass at the $\sqrt s = 7$~TeV LHC. The plots demonstrate various flavor-universal scenarios for the quark charge assignments and within each plot three benchmark values of the gauge mixing angle (c.f. \eqref{MC}) are shown, where we have defined: $r_{L,R} \equiv \frac{g_{L,R}}{g_{s}}$ (c.f. \eqref{Lferm}). The NLO corrections significantly reduce the dependence on the factorization scale $\mu_{F}$ (of the order of 2\% \cite{Chivukula:2011ng}), as evident from the thickness of the plotted curves. For comparison, the CMS experimental upper limits \cite{CMS} are superimposed, indicating a lower bound on the coloron mass in the TeV region.

\begin{figure}
\includegraphics[width=.33\textwidth]{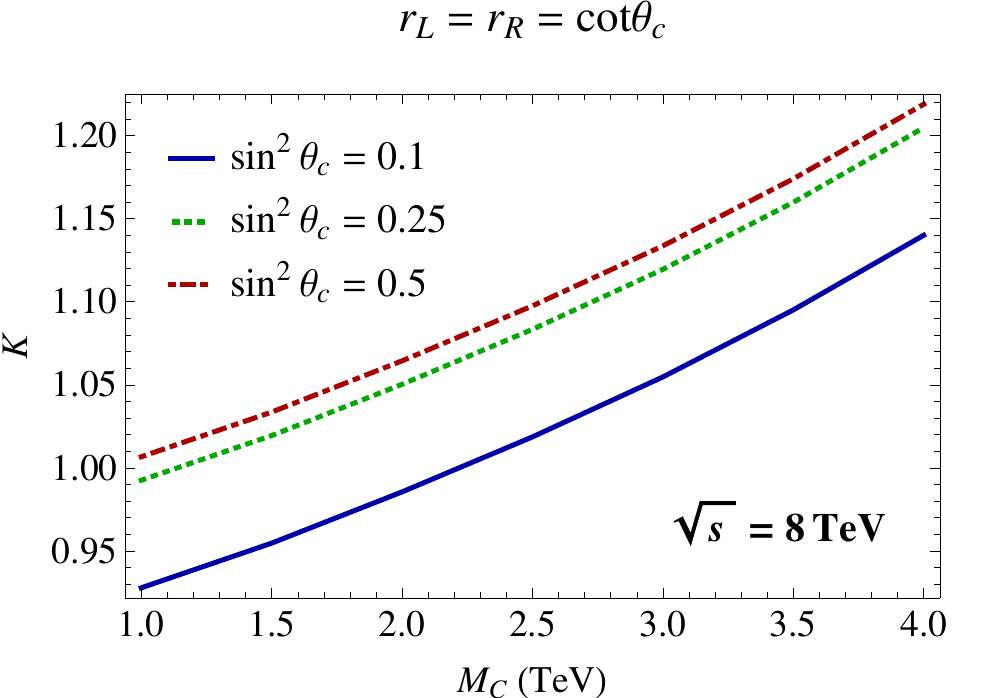}
\includegraphics[width=.33\textwidth]{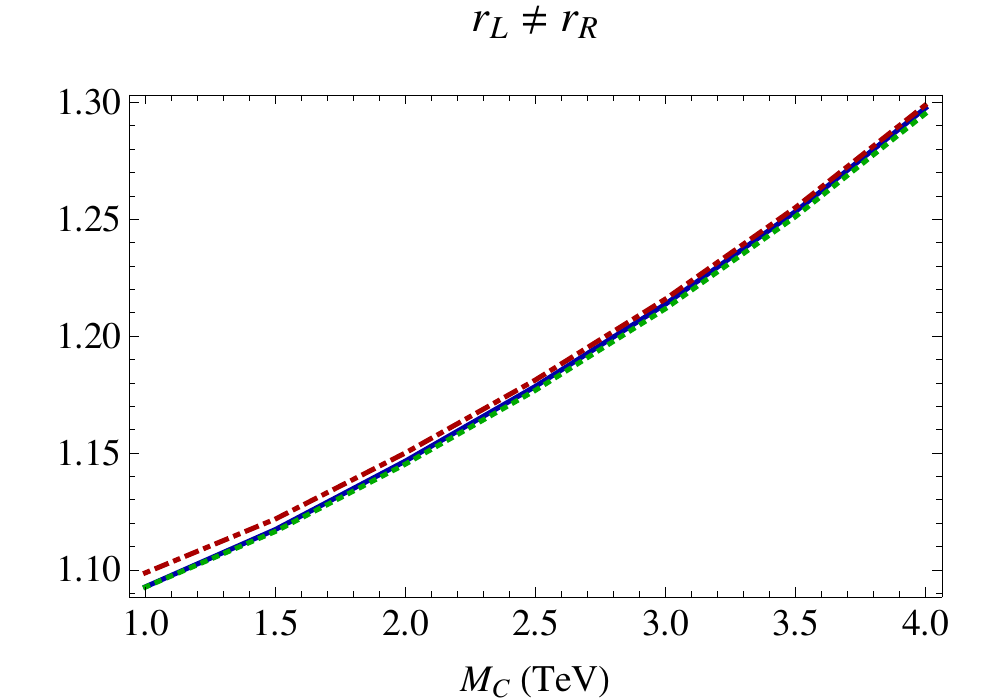}
\includegraphics[width=.33\textwidth]{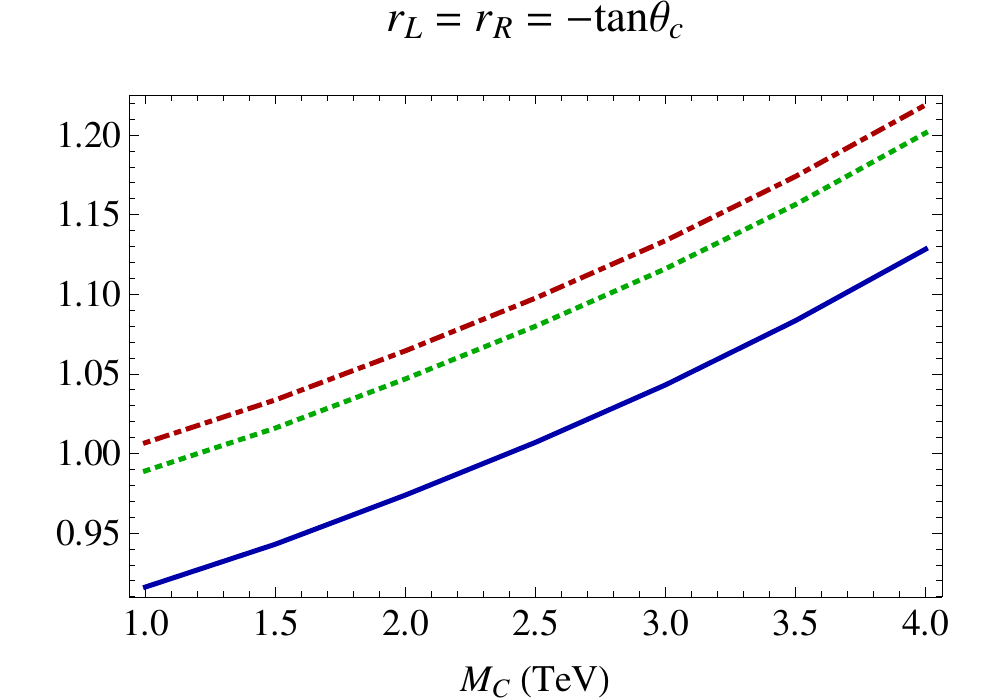}
\includegraphics[width=.33\textwidth]{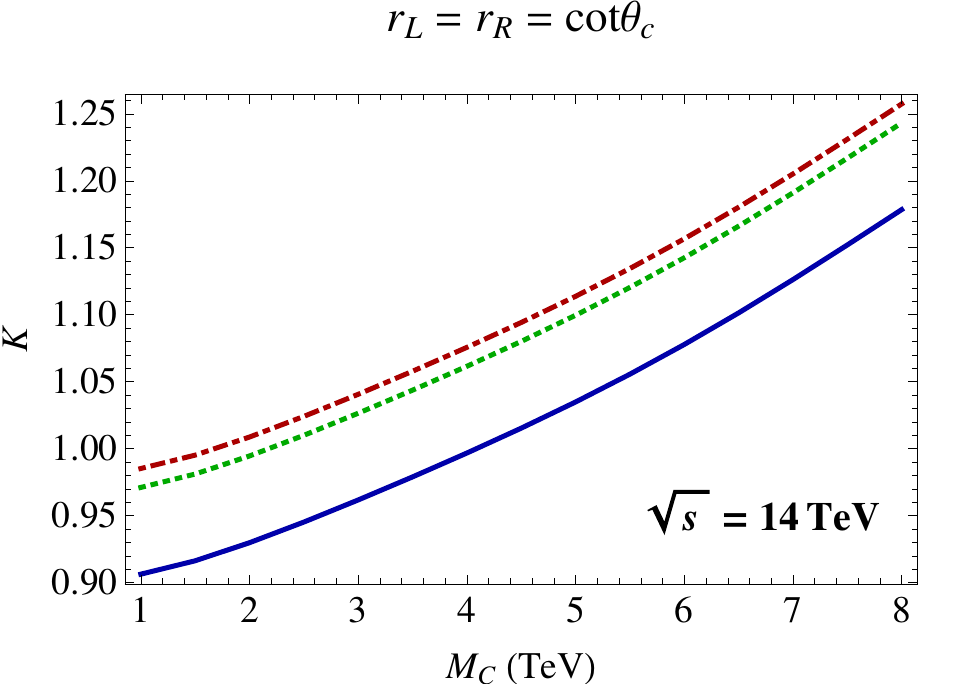}
\includegraphics[width=.33\textwidth]{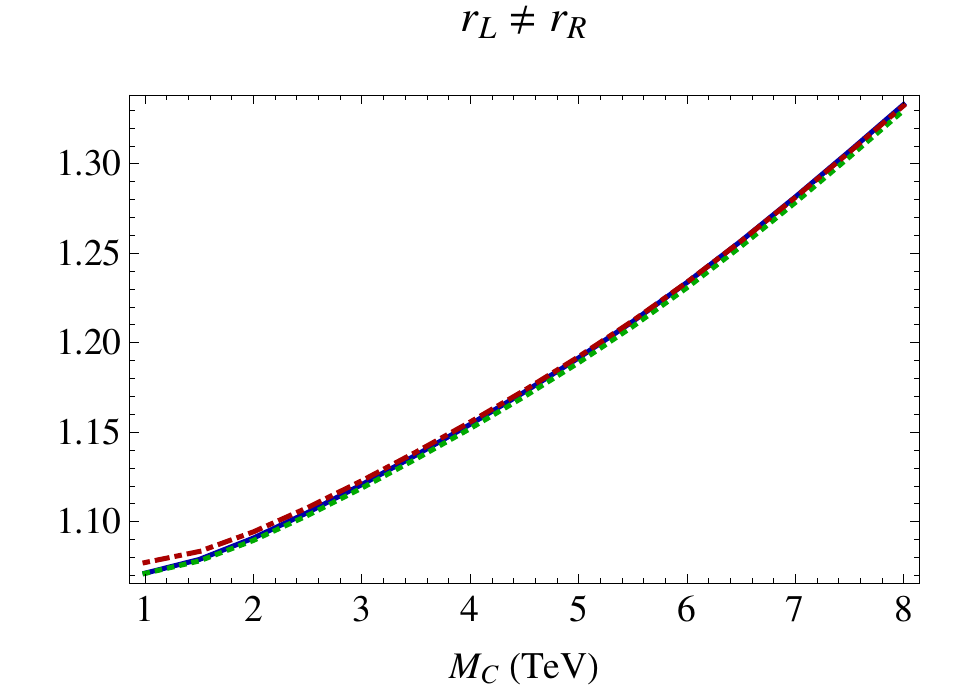}
\includegraphics[width=.33\textwidth]{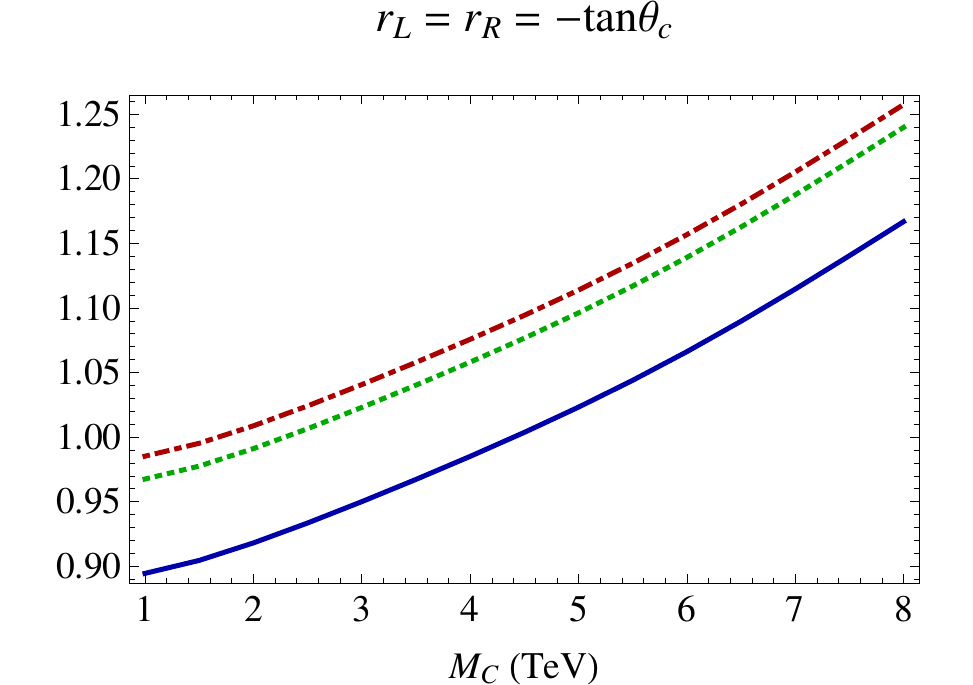}
\caption{The $K$-factor values for the LHC $\sqrt{s}=8$~TeV (top row) and $\sqrt{s}=14$~TeV (bottom row). Three possible flavor-universal scenarios for the quark charge assignments are considered, with different benchmark values of the gauge mixing angle within each plot ($\mu_{F} = M_{C}$). Note that the NLO corrections can be as large as 30\%. (plots taken from \cite{Chivukula:2013xla})}
\label{Kfac}
\end{figure}

The effects of the NLO corrections are most clearly represented by the $K$-factor, defined as
\begin{equation}\label{Kf}
K \equiv \frac{\sigma^{\text{NLO}}}{\sigma^{\text{LO}}} \ .
\end{equation}
The $K$-factor plots for various quark charge assignments and for different values of the gauge mixing angle are depicted in Fig.~\ref{Kfac} as a function of the coloron mass for the $\sqrt s = 8, 14$~TeV LHC. It is evident, once more, that the NLO corrections are significant, with the $K$-factor as large as 30\%, emphasizing the grave necessity of the higher-order calculations in reducing the theoretical uncertainties for a more accurate comparison with the experimental data. Such large NLO corrections predominantly originate from the initial state gluons in the Compton production process (a gluon and a (anti)quark in the initial state with a coloron and a (anti)quark in the final state), due to the abundance of gluons in a high center-of-mass energy hadron collider.

\begin{figure}
\begin{center}
\includegraphics[width=.5\textwidth]{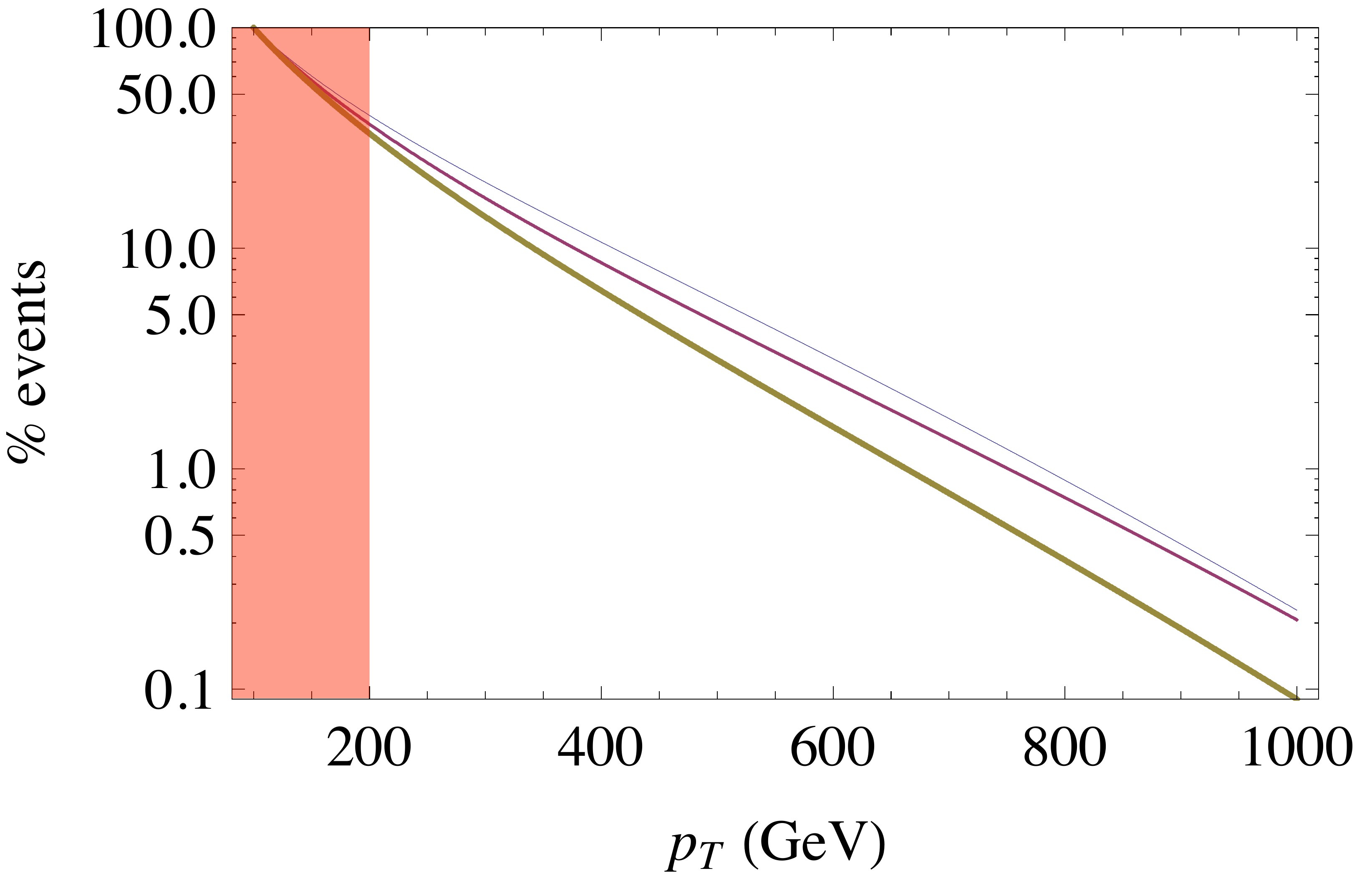}
\caption{Fraction of colorons produced with $p_T \ge p_{Tmin}$, as a function of $p_{Tmin}$, for the benchmark vector-like case $r_L=r_R=-\tan\theta_c$ and $\sin^2\theta_c = 0.05$. Three curves corresponding to various coloron masses are shown: $M_C=1.2$ (thin blue), 2.0 (medium purple), and 3.0 TeV (thick green). Within this mass range, about 30\% of the colorons are produced with $p_T \ge 200$~GeV. Below this $p_{T}$ value, the terms proportional to $\log(M^2_C/p^2_{Tmin})$ become large, and the fixed-order calculation is unreliable (red shaded region). (plot taken from \cite{Chivukula:2011ng})}
\label{pT}
\end{center}
\end{figure}

Finally, Fig.~\ref{pT} displays the fraction of colorons produced with a transverse momentum, $p_T$, greater than $p_{Tmin}$, as a function of $p_{Tmin}$, for a benchmark scenario and several coloron masses. Since the computation of the transverse momentum inherently starts at NLO, our calculation represents the LO results for this quantity. One observes that of the order of 30\% of the colorons can be produced with $p_T \ge 200$~GeV, making their collider searches very promising.

\section{Conclusion}\label{concl}

The renormalizable coloron model constitutes the minimal extension of the SM color sector, with the spontaneous symmetry breaking of the extended color gauge group facilitated by the renormalizable operators. It predicts the existence of the beyond the SM massive color-octet gauge bosons, colored and uncolored scalar degrees of freedom, as well as potential spectator fermions necessary for anomaly-cancelation purposes. In addition, keeping the ordinary chiral quark charge assignments under the extended color gauge group in their most general form, the model serves as a general simplified framework, (effectively) encompassing a large class of models propsed within the community.

The framework is well-constrained and highly predictive, and the hadron colliders form an ideal environment for exploring its properties; in particular, its free parameter space can be thoroughly probed by the LHC and the next generation hadron colliders. To this end, the NLO computations have been shown to be indispensable in order to significantly reduce the theoretical uncertainties, enabling more accurate comparisons with the experimental data.

\section*{Acknowledgment}

I am grateful to Sekhar Chivukula for comments on the draft.


\begin{thebibliography}{99}

\bibitem{Chivukula:2013xka} 
  R.~S.~Chivukula, A.~Farzinnia, J.~Ren and E.~H.~Simmons,
  Phys.\ Rev.\ D {\bf 88}, 075020 (2013)
  [arXiv:1307.1064 [hep-ph]].
  
\bibitem{Chivukula:2014rka} 
  R.~S.~Chivukula, E.~H.~Simmons, A.~Farzinnia and J.~Ren,
  Phys.\ Rev.\ D {\bf 90}, 015013 (2014)
  [arXiv:1404.6590 [hep-ph]].
  
\bibitem{Chivukula:2015kua} 
  R.~S.~Chivukula, A.~Farzinnia and E.~H.~Simmons,
  Phys.\ Rev.\ D {\bf 92}, no. 5, 055002 (2015)
  [arXiv:1504.03012 [hep-ph]].
  
\bibitem{Chivukula:2011ng} 
  R.~S.~Chivukula, A.~Farzinnia, E.~H.~Simmons and R.~Foadi,
  Phys.\ Rev.\ D {\bf 85}, 054005 (2012)
  [arXiv:1111.7261 [hep-ph]];
  
\bibitem{Chivukula:2013xla} 
 R.~S.~Chivukula, A.~Farzinnia, J.~Ren and E.~H.~Simmons,
  Phys.\ Rev.\ D {\bf 87}, no. 9, 094011 (2013)
  [arXiv:1303.1120 [hep-ph]].
  
\bibitem{Hill:1991at} 
  C.~T.~Hill,
  Phys.\ Lett.\ B {\bf 266}, 419 (1991).

\bibitem{Chivukula:1996yr} 
  R.~S.~Chivukula, A.~G.~Cohen and E.~H.~Simmons,
  Phys.\ Lett.\ B {\bf 380}, 92 (1996)
  [hep-ph/9603311].
  
\bibitem{axigluon} 
  P.~H.~Frampton and S.~L.~Glashow,
  Phys.\ Lett.\ B {\bf 190}, 157 (1987);
 J.~Bagger, C.~Schmidt and S.~King,
  Phys.\ Rev.\ D {\bf 37}, 1188 (1988).
 
\bibitem{Martynov:2009en} 
  M.~V.~Martynov and A.~D.~Smirnov,
  Mod.\ Phys.\ Lett.\ A {\bf 24}, 1897 (2009)
  [arXiv:0906.4525 [hep-ph]].
  
\bibitem{Frampton:2009rk} 
  P.~H.~Frampton, J.~Shu and K.~Wang,
  Phys.\ Lett.\ B {\bf 683}, 294 (2010)
  [arXiv:0911.2955 [hep-ph]].  

\bibitem{Top-Coloron}
  R.~S.~Chivukula, E.~H.~Simmons and N.~Vignaroli,
  Phys.\ Rev.\ D {\bf 87}, no. 7, 075002 (2013)
  [arXiv:1302.1069 [hep-ph]];
  R.~S.~Chivukula, E.~H.~Simmons and N.~Vignaroli,
  Phys.\ Rev.\ D {\bf 88}, 034006 (2013)
  [arXiv:1306.2248 [hep-ph]].
  
\bibitem{Chivukula:1995dt}
See R.~S.~Chivukula, R.~Rosenfeld, E.~H.~Simmons and J.~Terning,
  In *Barklow, T.L. (ed.) et al.: Electroweak symmetry breaking and new physics at the TeV scale* 352-382
  [hep-ph/9503202], and references therein.
  
\bibitem{KKg}
  H.~Davoudiasl, J.~L.~Hewett, T.~G.~Rizzo,
  Phys.\ Rev.\  {\bf D63}, 075004 (2001)
  [hep-ph/0006041]; 
  B.~Lillie, L.~Randall and L.~-T.~Wang,
  JHEP {\bf 0709}, 074 (2007)
  [hep-ph/0701166].
  
  \bibitem{recom}
  C.~T.~Hill and S.~J.~Parke,
  Phys.\ Rev.\ D {\bf 49}, 4454 (1994)
  [hep-ph/9312324];
  D.~A.~Dicus, B.~Dutta and S.~Nandi,
  Phys.\ Rev.\ D {\bf 51}, 6085 (1995)
  [hep-ph/9412370];
  Y.~Bai and B.~A.~Dobrescu,
  JHEP {\bf 1107}, 100 (2011)
  [arXiv:1012.5814 [hep-ph]].
  
  \bibitem{LHCHiggs}
G. Aad \textit{et al.} (ATLAS),
Phys.Lett. \textbf{B716}, 1 (2012),
arXiv:1207.7214 [hep-ex];
S. Chatrchyan \textit{et al.} (CMS),
Phys.Lett. \textbf{B716}, 30 (2012),
arXiv:1207.7235 [hep-ex].

\bibitem{Aaltonen:2013hya} 
  T.~Aaltonen {\it et al.}  [CDF Collaboration],
  Phys.\ Rev.\ Lett.\  {\bf 111}, no. 3, 031802 (2013)
  [arXiv:1303.2699 [hep-ex]].

\bibitem{ColoronLim}   
  E.~H.~Simmons,
  Phys.\ Rev.\ D {\bf 55}, 1678 (1997)
  [hep-ph/9608269];
  I.~Bertram and E.~H.~Simmons,
  Phys.\ Lett.\ B {\bf 443}, 347 (1998)
  [hep-ph/9809472];
  G.~Aad {\it et al.}  [ATLAS Collaboration],
  JHEP {\bf 1301}, 029 (2013)
  [arXiv:1210.1718 [hep-ex]];
  [ATLAS Collaboration],
  ATLAS-CONF-2012-148;
 S.~Chatrchyan {\it et al.}  [CMS Collaboration],
  Phys.\ Rev.\ D {\bf 87}, no. 11, 114015 (2013)
  [arXiv:1302.4794 [hep-ex]];
  [CMS Collaboration],
  CMS-PAS-EXO-12-059.
  G.~Aad {\it et al.}  [ATLAS Collaboration],
  Phys.\ Rev.\ D {\bf 91}, no. 5, 052007 (2015)
  [arXiv:1407.1376 [hep-ex]].
  V.~Khachatryan {\it et al.}  [CMS Collaboration],
  arXiv:1501.04198 [hep-ex].
  
  \bibitem{Spect} 
  CMS Collaboration [CMS Collaboration],
  CMS-PAS-B2G-12-015;
  [ATLAS Collaboration],
  ATLAS-CONF-2013-018;
For a general review of heavy electroweak vector fermions, see e.g.
S.~A.~R.~Ellis, R.~M.~Godbole, S.~Gopalakrishna and J.~D.~Wells,
  JHEP {\bf 1409}, 130 (2014)
  [arXiv:1404.4398 [hep-ph]].
    G.~Aad {\it et al.}  [ATLAS Collaboration],
  JHEP {\bf 1411}, 104 (2014)
  [arXiv:1409.5500 [hep-ex]].
  
  \bibitem{HeavyH}
  [CMS Collaboration],
  CMS-PAS-HIG-13-002; CMS-PAS-HIG-13-003;
  [ATLAS Collaboration],
  ATLAS-CONF-2013-013; ATLAS-CONF-2013-030.
  
\bibitem{ATLASHeavyS}
  [ATLAS Collaboration],
 ATL-PHYS-PUB-2013-016.

\bibitem{ATLASHproj}
  [ATLAS Collaboration],
  ATL-PHYS-PUB-2013-014.

\bibitem{CMS} 
  S.~Chatrchyan {\it et al.}  [CMS Collaboration],
  Phys.\ Lett.\ B {\bf 704}, 123 (2011)
  [arXiv:1107.4771 [hep-ex]].

\end{thebibliography}
\end{document}